\documentclass[12pt]{article}

\usepackage{latexsym}

\newcommand{\sig}[1]{\ensuremath{\sigma_{#1}}}
\newcommand{\gam}[1]{\ensuremath{\gamma_{#1}}}
\newcommand{\ugam}[1]{\ensuremath{\gamma^{#1}}}
\newcommand{\I}{\ensuremath{\mathcal{I}}}
\newcommand{\scalar}[1]{\ensuremath{\langle#1\rangle_{s}}}
\newcommand{\multi}[2]{\ensuremath{\langle#1\rangle_{#2}}}

\begin{document}

%\baselineskip=30

%------------------- BEGIN TITLE ----------------------------------
\begin{flushleft}
\textbf{DYNAMIC AND GEOMETRIC PHASE
         FORMULAS IN THE HESTENES-DIRAC THEORY}\newline

\begin{quote}\begin{quote}
        \textbf{David W. Dreisigmeyer}\newline\newline
        \textit{Department of Electrical and Computer Engineering}\\
        \textit{Colorado State University, Fort Collins, CO 80523}\\
        \textit{email:davidd@engr.colostate.edu}\newline\newline
        \textbf{Richard Clawson}\newline\newline
        \textit{Department of Physics and Astronomy}\\
        \textit{Arizona State University, Tempe, AZ 85287}\\
        \textit{email:richard.clawson@asu.edu}\newline\newline
        \textbf{R. Eykholt}\newline\newline
        \textit{Department of Physics}\\
        \textit{Colorado State University, Fort Collins, CO 80523}\\
        \textit{email:eykholt@lamar.colostate.edu}\newline\newline
        \textbf{Peter M. Young}\newline\newline
        \textit{Department of Electrical and Computer Engineering}\\
        \textit{Colorado State University, Fort Collins, CO 80523}\\
        \textit{email:pmy@engr.colostate.edu}\newline\newline\newline
\end{quote}\end{quote}
\end{flushleft}

%---------------BEGIN ABSTRACT-------------------------------------
\noindent We examine the dynamic and geometric phases of the
electron in quantum mechanics using Hestenes' spacetime algebra
formalism. First the standard dynamic phase formula is translated
into the spacetime algebra.  We then define new formulas for the
dynamic and geometric phases that can be used in Hestenes'
formalism.

%----------------BEGIN INTRODUCTION---------------------------------------
%\section{INTRODUCTION}
%\label{Introduction}
\noindent\textbf{1.  INTRODUCTION}\newline

The geometric phase is now a standard topic in quantum mechanics
[\ref{Sak}].  To give an idea of what the geometric phase is, let
us assume we have a parameter dependent Hamiltonian
$H(\mathbf{P}(t))$ where the parameters are time dependent [e.g.,
a time dependent magnetic field].  As a quantum system evolves
under the Hamiltonian, it will pick up a total phase factor.  We
can decompose this phase into two parts.  The first part is the
dynamic phase which reflects how long the system has been
evolving.  The second part, the geometric phase part of the total
phase, depends only on the path that $\mathbf{P}(t)$ takes in its
evolution.  That is, the geometric phase reflects the geometry of
$\mathbf{P}$'s evolution in parameter space.  An important point
is that the geometric phase does not depend on how fast
$\mathbf{P}(t)$ traverses its path, but depends only on the path
itself. The article by Mukunda and Simon [\ref{MuS1}] offers
perhaps the most easily understood comprehensive introduction to
the theory.

Here we examine the dynamic and geometric phases in Hestenes'
spacetime algebra formalism [\ref{HSTA}].  This is an alternative
mathematical treatment of the Dirac equation.  So, the system we
have in mind is a relativistic spin-1/2 particle evolving under a
time dependent Hamiltonian. The time dependence of the Hamiltonian
will come from a time varying electromagnetic field.

The paper is organized as follows. In Section 2, we review the
spacetime algebra formalism.  The dynamic and geometric phase
formulas are derived in Section 3.  A discussion of the results
follows in Section 4.\newline\newline

%-------------BEGIN GEOMETRIC ALGEBRA--------------------------------------
%\section{SPACETIME ALGEBRA}
%\label{Geometric}
\noindent\textbf{2. SPACETIME ALGEBRA}\newline

We review the spacetime algebra formalism that will be needed in
the sequel.  The reader is referred to [\ref{HSTA}] for a more
detailed account.  We will let $\Psi$ and $\Phi$ denote the
traditional [i.e., complex] wavefunctions, and $\psi$ and $\phi$
denote their spacetime algebra representations.

Hestenes' spacetime algebra is the geometric [or Clifford] algebra
of flat Minkowski spacetime.  Before looking at this algebra, we
will examine a simpler geometric algebra, the Pauli algebra.
First, we will take as an axiom that the \textit{vector} [or
\textit{Clifford}] \textit{product} of a vector $\mathbf{v}$ with
itself is given by $\mathbf{vv} := \mathbf{v} \cdot \mathbf{v}$,
where $\mathbf{v}\cdot\mathbf{v}$ is the usual \textit{inner
product}. For our example we will have three vectors \sig{1},
\sig{2} and \sig{3}.  Let $\mathbf{v} = \sig{1} + \sig{2}$ so that
\begin{eqnarray}
\mathbf{vv} & = & (\sig{1} + \sig{2})(\sig{1} + \sig{2})
                                \nonumber           \\
            & = & \sig{1}\sig{1} + \sig{1}\sig{2} + \sig{2}\sig{1}
                    + \sig{2}\sig{2}    \label{Product1}
\end{eqnarray}
and
\begin{eqnarray}
\mathbf{v}\cdot\mathbf{v} & = & (\sig{1} + \sig{2})\cdot (\sig{1}
                + \sig{2})      \nonumber       \\
            & = & \sig{1}\cdot\sig{1} + 2\sig{1}\cdot\sig{2} +
                \sig{2}\cdot\sig{2}.     \label{Product2}
\end{eqnarray}
Since $\mathbf{vv} = \mathbf{v}\cdot\mathbf{v}$, and using
$\sig{1}\sig{1} = \sig{1}\cdot\sig{1}$ and $\sig{2}\sig{2} =
\sig{2}\cdot\sig{2}$, we see from (\ref{Product1}) and
(\ref{Product2}) that
\begin{eqnarray}\label{InnerProduct1}
\sig{1}\cdot\sig{2} & = & \frac{1}{2}(\sig{1}\sig{2} +
            \sig{2}\sig{1}).
\end{eqnarray}
Equation (\ref{InnerProduct1}) allows us to define the inner
product in terms of the vector product
\begin{eqnarray}\label{InnerProduct2}
\sig{i}\cdot\sig{j} & = & \frac{1}{2}(\sig{i}\sig{j} +
            \sig{j}\sig{i}).
\end{eqnarray}
Because $\sig{i}\cdot\sig{j}$ is a scalar, we can form a metric
tensor by defining $\eta_{ij} := \sig{i}\cdot\sig{j}$, where $i,j
= 1,2,3$.  Then (\ref{InnerProduct2}) becomes
\begin{eqnarray}\label{MetricTensor}
\sig{i}\cdot\sig{j} & = & \eta_{ij}.
\end{eqnarray}

Now notice that
\begin{eqnarray}\label{OuterProduct1}
\sig{i}\sig{j} & = & \frac{1}{2}(\sig{i}\sig{j} + \sig{j}\sig{i})
        + \frac{1}{2}(\sig{i}\sig{j} - \sig{j}\sig{i}).
\end{eqnarray}
The first part of (\ref{OuterProduct1}) is the inner product of
\sig{i} and \sig{j}.  The second part of (\ref{OuterProduct1}) we
will call the \textit{outer product} of \sig{i} and \sig{j}, and
denote this by $\sig{i}\wedge\sig{j} := 1/2(\sig{i}\sig{j} -
\sig{j}\sig{i})$.  So (\ref{OuterProduct1}) can be rewritten as
\begin{eqnarray}\label{OuterProduct2}
\sig{i}\sig{j} & = & \sig{i}\cdot\sig{j} + \sig{i}\wedge\sig{j}.
\end{eqnarray}
What happens if $\sig{i} = \lambda\sig{j}$, where $\lambda$ is a
scalar?  Then,
\begin{eqnarray}
\sig{i}\wedge\sig{j} & = & \frac{1}{2}(\sig{i}\sig{j} -
            \sig{j}\sig{i})         \\
        & = & \frac{\lambda}{2}(\sig{j}\sig{j} - \sig{j}\sig{j})
                        \nonumber           \\
        & = & 0.     \nonumber
\end{eqnarray}
So, if \sig{i} and \sig{j} are not a multiple of each other [i.e.,
if they are not collinear], our vector product will have a nonzero
outer product.

The geometry of geometric algebra is seen by how we interpret all
of this.  As usual, we consider the inner product
$\sig{i}\cdot\sig{j}$ as being the projection of $\sig{i}$ onto
$\sig{j}$.  But what of $\sig{i}\wedge\sig{j}$?  This we let
represent the area between the vectors \sig{i} and \sig{j} in a
plane that contains both vectors.

Let us now make some refinements.  First, instead of representing
a [real] scalar as simply $\lambda$, we will denote it by $\lambda
1$.  Since scalars will commute with vectors [e.g.,
$\lambda\sig{1} = \sig{1}\lambda$], we will have
\begin{eqnarray}\label{Scalar}
1 \sig{i} & = & \sig{i}1.
\end{eqnarray}
Now, what if our vectors are orthonormal so that
\begin{eqnarray}\label{Orthonormal}
\sig{i}\cdot\sig{j} & = & \delta_{ij}?
\end{eqnarray}
Then, if $i \neq j$,
\begin{eqnarray}\label{OrthoProduct}
\sig{i}\sig{j} & = & \sig{i}\cdot\sig{j} + \sig{i}\wedge\sig{j}
                        \nonumber           \\
                & = & \sig{i}\wedge\sig{j}  \nonumber   \\
                & = & -\sig{j}\wedge\sig{i} \nonumber   \\
                & = & -\sig{j}\sig{i}.
\end{eqnarray}
So, if \sig{i} and \sig{j} are different orthonormal vectors, the
outer product $\sig{i}\wedge\sig{j}$ is not simply an area.
Rather, it is a directed area since $\sig{i}\sig{j} =
-\sig{j}\sig{i}$.  So it depends on whether we go ``clockwise'' or
``anti-clockwise'' over the area.  This is a generalization of the
concept of a vector's direction.  To formalize this new concept,
call the elements \sig{i}\sig{j} \textit{bivectors}.  Since $i,j =
1,2,3$, there are three independent bivectors given by
\sig{1}\sig{2}, \sig{1}\sig{3} and \sig{2}\sig{3} [note that, for
example, $\sig{1}\sig{2} = -\sig{2}\sig{1}$ and $\sig{3}\sig{3} =
\sig{3}\cdot\sig{3}$].  A further generalization is to have a
directed volume element, or a \textit{trivector}, $\I :=
\sig{1}\sig{2}\sig{3}$.  The trivector satisfies
\begin{eqnarray}\label{Isquared}
\mathcal{I}^{2} & = & (\sig{1} \sig{2} \sig{3}) (\sig{1} \sig{2}
                        \sig{3})        \nonumber       \\
                & = & -\sig{1}\sig{1} \sig{2}\sig{2}
                    \sig{3}\sig{3}      \nonumber       \\
                & = & -1.
\end{eqnarray}

So, by starting with three vectors \sig{1}, \sig{2} and \sig{3}
and a vector product, we are naturally lead to new elements $1$,
$\mathcal{I}$ and the bivectors.  That is, given three vectors, we
can imagine, in addition to the vectors themselves, a scalar
component, a directed volume and directed areas.  So our basis for
this construction is given by
$\{1,\sig{i},\sig{j}\sig{k},\sig{1}\sig{2}\sig{3}\}$, where $i =
1,2,3$ and $1 \leq j < k \leq 3$.  Notice that
\begin{eqnarray}
\I\sig{1} & = & \sig{1}\sig{2}\sig{3}\sig{1}    \nonumber   \\
        & = & \sig{1}\sig{1}\sig{2}\sig{3}      \nonumber   \\
        & = & \sig{2}\sig{3}
\end{eqnarray}
and that $\I\sig{2} = -\sig{1}\sig{3}$ and $\I\sig{3} =
\sig{1}\sig{2}$.  So we can also use $\{1,\sig{i},\I\sig{i},\I\}$,
$i = 1,2,3$, as our basis.  We prefer to use this latter basis.

It is possible to represent our vectors as matrices.  For our
example, the familiar Pauli spin matrices
\begin{eqnarray}
\begin{array}{ccc}
\widehat{\sigma}_{1} = \left[\begin{array}{cc}
            0 & 1   \\
            1 & 0
            \end{array}\right]          &
\widehat{\sigma}_{2} = \left[\begin{array}{cc}
            0 & -i  \\
            i & 0
            \end{array}\right]          &
\widehat{\sigma}_{3} = \left[\begin{array}{cc}
            1 & 0      \\
            0 & -1
            \end{array}\right],
\end{array}
\end{eqnarray}
where $i = \sqrt{-1}$, are one possible representation.  [In a
matrix representation we let $1$ be the identity matrix.]  While
it is nice to have an explicit matrix representation for our
vectors, we must stress that they are to be treated as vectors.
That means that under a spatial rotation, the basis vectors will
change.  For example, if we rotate in the \sig{1}\sig{2}-plane by
an angle $\varphi$, a vector will transform as
\begin{eqnarray}\label{VectorTransf}
\mathbf{w} & \rightarrow & \mathbf{w}^{'} = U\mathbf{w}U^{-1},
\end{eqnarray}
where $U = \exp[-\I\sig{3}\varphi/2]$.  So, if $\varphi = \pi/2$,
we expect that $\sig{1} \rightarrow \sig{2}$, $\sig{2} \rightarrow
-\sig{1}$ and $\sig{3} \rightarrow \sig{3}$.  This is easily
checked since, e.g.,
\begin{eqnarray}
\exp\left[-\I\sig{3}\frac{\pi}{4}\right]\sig{2}\exp\left[\I\sig{3}\frac{\pi}{4}\right]
        & = & \exp\left[-\I\sig{3}\frac{\pi}{2}\right]\sig{2}    \nonumber   \\
        & = & \left[\cos\left(\frac{\pi}{2}\right) - \I\sig{3}
            \sin\left(\frac{\pi}{2}\right)\right]\sig{2}
            \nonumber       \\
        & = & -\I\sig{3}\sig{2}     \nonumber       \\
        & = & -\sig{1},
\end{eqnarray}
where we used the facts that $\I\sig{i} = \sig{i}\I$ in the first
line, and that $(\I\sig{i})^{2} = -1$ in the second. Equation
(\ref{VectorTransf}) also holds for a general rotation using the
Euler angles $(\varphi, \theta, \chi)$. In this case $U$ has the
form [\ref{GLD1}]
\begin{eqnarray}\label{Urotation}
U & = & \exp\left[-\I\sig{3}\frac{\varphi}{2}\right]
    \exp\left[-\I\sig{2}\frac{\theta}{2}\right]
    \exp\left[-\I\sig{3}\frac{\chi}{2}\right].
\end{eqnarray}

So far we have only considered scalars, vectors, etc.  We can also
have more general objects, called \textit{multivectors}, which are
of the form
\begin{eqnarray}\label{Multivector}
C & = & c^{s} + (c_{1}^{v}\sig{1} + c_{2}^{v}\sig{2} +
        c_{3}^{v}\sig{3}) + \I(c_{1}^{b}\sig{1} + c_{2}^{b}\sig{2}
        + c_{3}^{b}\sig{3}) + c^{t}\I,
\end{eqnarray}
where $C$ has scalar [$c^{s}$], vector [$c_{i}^{v}$], bivector
[$c_{i}^{b}$] and trivector [$c^{t}$] parts.  Let us introduce two
operations on $C$ that will prove useful in the sequel.  The first
is to let \scalar{C} denote the scalar part of
(\ref{Multivector}), so $\scalar{C} = c^{s}$.  For higher parts of
$C$, we will let $\multi{C}{i}$ be the part of $C$ that can be
minimally expressed using $i$ vectors.  So, $\multi{C}{1} =
c_{1}^{v}\sig{1} + c_{2}^{v}\sig{2} + c_{3}^{v}\sig{3}$,
$\multi{C}{2} = \I(c_{1}^{b}\sig{1} + c_{2}^{b}\sig{2} +
c_{3}^{b}\sig{3})$ and $\multi{C}{3} = c^{t}\I$. The second
operation is the \textit{reversion} of $C$, written as
$\widetilde{C}$.  By this we mean the reversal of all the vector
products in $C$.  So, for example, $\widetilde{\sig{1}\sig{2}} =
\sig{2}\sig{1}$.  In general, for two multivectors $A$ and $B$, we
have $\widetilde{AB} = \widetilde{B}\widetilde{A}$.  For
(\ref{Multivector}) notice that $\widetilde{\I} = -\I$ and
$\widetilde{\I\sig{i}} = -\sig{i}\I = -\I\sig{i}$ [because
$\I\sig{i} = \sig{i}\I$].  Then
\begin{eqnarray}\label{Reversion}
\widetilde{C} & = & c^{s} + (c_{1}^{v}\sig{1} +
        c_{2}^{v}\sig{2} +
        c_{3}^{v}\sig{3}) - \I(c_{1}^{b}\sig{1} + c_{2}^{b}\sig{2}
        + c_{3}^{b}\sig{3}) - c^{t}\I,
\end{eqnarray}
We can use the reversion operation to rewrite
(\ref{VectorTransf}).  From (\ref{Urotation}) we see that $U^{-1}
= \widetilde{U}$.  So (\ref{VectorTransf}) can be written as
\begin{eqnarray}\label{ReversionRotation}
C & \rightarrow & C^{'} = U C \widetilde{U},
\end{eqnarray}
where we now allow the rotation to act on a general multivector.

Now we will review Hestenes' spacetime algebra.  The spacetime
algebra is generated by four vectors $\{\gam{\mu}\}$, $\mu =
0,1,2,3$, that satisfy the Dirac algebra
\begin{eqnarray}\label{DiracAlgebra}
\gam{\mu}\cdot\gam{\nu} & = & g_{\mu\nu}    \\
        & = & \mbox{diag}(+\ -\ -\ -). \nonumber
\end{eqnarray}
The metric tensor $g_{\mu\nu}$ in (\ref{DiracAlgebra}) is the
usual one from special relativity.  The \gam{0} vector is the time
direction and the \gam{n}, $n=1,2,3$, vectors are the spatial
directions, in an observer's frame of reference.  We can also
raise the indices on the \gam{\mu}'s by defining the vectors
\ugam{\nu} as those that satisfy the relation
$\ugam{\nu}\cdot\gam{\mu} = \delta^{\nu}_{\mu}$. [Notice that
$\gamma^{0} = \gam{0}$ and $\gamma^{n} =- \gam{n}$ satisfy this.]
One possible matrix representation of the \gam{\mu}'s are the
standard Dirac matrices.

The vectors $\gam{\mu}$ result in the basis
\begin{eqnarray}\label{DiracBasis}
\left\{1,\gam{\mu},(\sig{n},\I\sig{n}),\I\gam{\mu},\I\right\},
\end{eqnarray}
where $n = 1,2,3$ and $\sig{n} := \gam{n}\gam{0}$.  Also, $\I :=
\gam{0}\gam{1}\gam{2}\gam{3} = \sig{1}\sig{2}\sig{3}$,
$\I\gam{\mu} = -\gam{\mu}\I$ and now $\widetilde{\I} = \I$.  The
Pauli algebra we examined above is a sub-algebra of the Dirac
algebra. So the $1$, $\sig{n}$, $\I\sig{n}$ and $\I$ in
(\ref{DiracBasis}) satisfy the geometric algebra we looked at
using the $\sig{n}$'s as our vectors.  We can consider the Pauli
algebra as the algebra of space and the Dirac algebra as the
algebra of spacetime.  Note that in the Pauli algebra, the
$\sig{n}$'s are vectors while the $\I\sig{n}$'s are bivectors.  In
the Dirac algebra, both of these are bivectors.

In the spacetime algebra formalism, a proper, orthochronous
Lorentz transformation [i.e., a spacetime rotation that does not
reverse time or space] is represented by $R$.  Since every Lorentz
transformation can be decomposed into a pure boost and a pure
spatial rotation, $R$ will have the form $R = LU$, where $U$ is
the pure spatial rotation given in (\ref{Urotation}), and the pure
boost, $L$, has the general form
\begin{eqnarray}\label{LRotation}
L & = & \exp\left[-\frac{(b_{1}\sig{1} + b_{2}\sig{2} +
        b_{3}\sig{3})}{2}\right].
\end{eqnarray}
The spacetime generalization of (\ref{VectorTransf}) is given by
\begin{eqnarray}\label{SpaceTimeVectorTransf}
\mathbf{w} & \rightarrow & \mathbf{w}^{'} = R\mathbf{w}R^{-1}.
\end{eqnarray}
We have already seen that $U^{-1} = \widetilde{U}$ which still
holds in the Dirac algebra [because $\widetilde{\sigma}_{n} =
-\sig{n}$, $\widetilde{\I} = \I$ and $\I\sig{n} = \sig{n}\I$, so
$\widetilde{(\I\sig{n})} = -\I\sig{n}$]. Because
$\widetilde{\sigma}_{n} = -\sig{n}$, we also have that $L^{-1} =
\widetilde{L}$, so $\widetilde{R} = \widetilde{U}\widetilde{L}=
R^{-1}$.  As we did in (\ref{ReversionRotation}), we can write
(\ref{SpaceTimeVectorTransf}) as
\begin{eqnarray}\label{SpaceTimeReversionRotation}
C & \rightarrow & C^{'} = R C \widetilde{R},
\end{eqnarray}
where $C$ is a general spacetime multivector.

It can be shown [\ref{H75-1}] that a Dirac spinor, in the
spacetime algebra, has the general form
\begin{eqnarray}\label{DiracSpinor}
\psi & = & \left(\rho e^{\I\beta}\right)^{1/2}R,
\end{eqnarray}
where $\rho(x)$ and $\beta(x)$ are scalars and $R(x)$ is a
[proper, orthochronous] Lorentz transformation.  [Deriving
(\ref{DiracSpinor}) is easy though lengthy, so we will simply
accept this form for $\psi$ without proof.]  Note that $\rho$,
$\beta$ and $R$ depend on the spacetime point $x$.  That is, they
are locally defined parameters in the wavefunction $\psi$.  In
particular, $R$ is a local Lorentz transformation-- \textit{not} a
global one.  This observation leads to the local observables
theory favored by Hestenes [\ref{H73}]. Now, the $\rho$ term in
(\ref{DiracSpinor}) is interpreted as a proper probability
density. A satisfactory interpretation of the $\beta$ parameter is
still lacking in Hestenes' theory, however. So Hestenes'
interpretation of the Dirac equation, while attractive, is not yet
complete. Using the form for $R$ above, (\ref{DiracSpinor}) can be
rewritten as
\begin{eqnarray}
\psi & = & \left(\rho e^{\I\beta}\right)^{1/2}LU
                \label{DiracSpinor1}        \\
    & = &\left(\rho e^{\I\beta}\right)^{1/2}LU_{0}
            e^{-\I\sig{3}\chi/2},
                \label{DiracSpinor2}
\end{eqnarray}
where, from (\ref{Urotation}),
\begin{eqnarray}\label{U0}
U_{0} & := & e^{-\I\sig{3}\varphi/2}e^{-\I\sig{2}\theta/2}.
\end{eqnarray}

Multiplying $\Psi$ in the standard Dirac theory by a phase factor
$\exp[i\alpha]$ corresponds, in the Hestenes-Dirac theory, to
\begin{eqnarray}\label{Multiplybyi}
\psi & \rightarrow & \psi^{'} = \psi e^{\I\sig{3}\alpha}.
\end{eqnarray}
Equation (\ref{Multiplybyi}) simply expresses a correspondence
between two different mathematical formalism, the standard Dirac
theory and Hestenes' theory.  It is useful to note two things
though.  First, $\I\sig{3}$, in Hestenes' formalism, corresponds
to the complex number $i$ in the standard Dirac theory.  Also,
given (\ref{DiracSpinor2}), multiplication by a phase factor in
the standard Dirac theory corresponds to a change in the Euler
angle $\chi$ in the spacetime algebra formalism.  This allows us
to identify $-\chi/2$ as the phase in Hestenes' theory.  [Note the
minus sign here.  In (\ref{Multiplybyi}) we \textit{added} a new
phase $\alpha$ to the original phase.  The minus sign follows from
(\ref{DiracSpinor2}).] Another lengthy, but easy, derivation shows
that the Hestenes form of the Dirac equation for a spin-$1/2$
particle of charge $e$ and mass $m$ is given by [with $\hbar = 1$
and $c = 1$]
\begin{eqnarray}\label{DiracEquation}
\Box \psi \I\sig{3} - eA\psi & = &
    m\psi \gam{0},
\end{eqnarray}
where $A := \ugam{\mu}A_{\mu}$ is the electromagnetic field vector
and $\Box := \ugam{\mu}\partial_{\mu}$. [The Einstein summation
convention is used here, so $\gamma^{\mu}A_{\mu} :=
\gamma^{0}A_{0} + \gamma^{1}A_{1} + \gamma^{2} A_{2} + \gamma^{3}
A_{3}$.]

In the standard Dirac theory, the probability density $\varrho$ in
an observer's frame of reference is associated with the time
component of the Dirac current $\Psi^{\dag}\Psi$, where
$\Psi^{\dag}$ is the Hermitian adjoint of the complex wavefunction
$\Psi$.  In order to translate this into Hestenes' formalism, we
need to find the spacetime algebra equivalent of
$\Psi^{\dag}\Phi$, where $\Phi$ is some other complex
wavefunction.  The spacetime algebra representation of
$\Psi^{\dag}$ is given by [\ref{DLG}]
\begin{eqnarray}\label{HermitianAdjoint}
\Psi^{\dag} & \leftrightarrow  \psi^{\dag} := \gam{0}
                \widetilde{\psi}\gam{0}.
\end{eqnarray}
Now, $\Psi^{\dag}\Phi = r + i c$ where $r$ and $c$ are real
scalars.  The real part of $\Psi^{\dag}\Phi$, denoted
$\Re(\Psi^{\dag}\Phi)$, is simply the real scalar $r$. This
corresponds to the scalar part of
$\phi\gam{0}\widetilde{\psi}\gam{0}$, the spacetime algebra
representation of $\Psi^{\dag}\Phi$.  Then $\Re(\Psi^{\dag}\Phi)
\leftrightarrow \scalar{\phi\gam{0}\widetilde{\psi}\gam{0}}$.  Now
recall that $i$ in the Dirac theory corresponds to $\I\sig{3}$ in
Hestenes' theory.  Then the imaginary part of $\Psi^{\dag}\Phi$,
written as $\Im(\Psi^{\dag}\Phi) = - \Re(i\Psi^{\dag}\Phi)$, goes
to $-\scalar{\phi\I\sig{3}\gam{0}\widetilde{\psi}\gam{0}}$ in the
spacetime algebra formalism.  So the spacetime algebra
representation of the complex probability  amplitude density of
standard quantum theory is given by [\ref{DLG}]
\begin{eqnarray}\label{ComplexAmplitude}
\Psi^{\dag}\Phi & = & \Re(\Psi^{\dag}\Phi) + i\Im(\Psi^{\dag}\Phi)
                        \nonumber           \\
            & \leftrightarrow &
            \scalar{\phi\gam{0}\widetilde{\psi}\gam{0}} -
            \I\sig{3}
            \scalar{\phi\I\sig{3}\gam{0}\widetilde{\psi}\gam{0}}.
\end{eqnarray}
Equation (\ref{ComplexAmplitude}) is equal to the probability
density $\varrho$ when $\Phi = \Psi$.

Notice the ubiquitous presence of \gam{0} in
(\ref{HermitianAdjoint}) and (\ref{ComplexAmplitude}).  Since
\gam{0} is the time vector in an observer's frame, we say that the
Hermitian adjoint $\Psi^{\dag}$ and the probability density
$\varrho$ are ``frame dependent''. That is, they depend on the
direction of \gam{0}, and, hence, on the observer's frame of
reference.

Results similar to (\ref{HermitianAdjoint}) and
(\ref{ComplexAmplitude}) hold for the Dirac adjoint
$\overline{\Psi}:=\Psi^{\dag}\widehat{\gamma}_{0}$, where
$\widehat{\gamma}_{0}$ is the standard Dirac matrix.  The
spacetime algebra representation of $\overline{\Psi}$ is
[\ref{DLG}]
\begin{eqnarray}\label{DiracAdjoint}
\overline{\Psi} & \leftrightarrow & \widetilde{\psi}.
\end{eqnarray}
Also,
\begin{eqnarray}\label{ComplexAmplitude2}
\overline{\Psi}\Phi & \leftrightarrow & \scalar{\phi
    \widetilde{\psi}} - \I\sig{3}\scalar{\phi\I\sig{3}
    \widetilde{\psi}}.
\end{eqnarray}
Since (\ref{DiracAdjoint}) and (\ref{ComplexAmplitude2}) do not
contain a $\gamma_{0}$ term, we say they are frame independent.

The $\psi$ in (\ref{DiracSpinor}) can also be rewritten in a form
like that in (\ref{Multivector}).  That is, $\psi$ can be
expressed as
\begin{eqnarray}\label{psiMultivector}
\psi & = & c^{s} + (c_{1}^{v}\sig{1} + c_{2}^{v}\sig{2} +
        c_{3}^{v}\sig{3}) + \I(c_{1}^{b}\sig{1} + c_{2}^{b}\sig{2}
        + c_{3}^{b}\sig{3}) + c^{t}\I,
\end{eqnarray}
where $\sig{n} = \gam{n}\gam{0}$ in the Dirac, versus Pauli,
algebra.  Now, all the $c$'s in (\ref{psiMultivector}) are real,
so it would appear that we have managed to remove the complex
structure of the Dirac theory by using the spacetime algebra. This
is incorrect, as (\ref{ComplexAmplitude}) shows.  The ``$1$'' and
``$\I\sig{3}$'' of Hestenes' theory correspond to the real and
imaginary parts, respectively, of the standard Dirac theory.  So,
even though we can express $\psi$ completely in terms of real
numbers, as in (\ref{psiMultivector}), the complex nature of the
Dirac theory still remains.  As explained in [\ref{Gspon2002}],
such a complex structure is inherent to any correct formulation of
Dirac's equation.

Lastly, we need to define a few terms that will be used latter.
[To rigorously justify these definitions will take us too far
afield. The reader can consult [\ref{H75-1}], and the references
therein, to find their physical justifications.  For now, they can
simply be taken as definitions.] The proper velocity vector $v =
v^{\mu}\gam{\mu}$ is defined by
\begin{eqnarray}\label{G7b}
v & := & R\gamma_{0}\widetilde{R}   \\
    & = & R_{0}\gamma_{0}\widetilde{R}_{0}, \label{vEq2}
\end{eqnarray}
where $R$ is from the wavefunction $\psi$ and $R_{0} := LU_{0}$.
That is, $v$ is given by the spacetime rotation of the time axis
\gam{0} determined locally by the wavefunction $\psi$.  We have
that $vv = 1$, so the velocity vector is normalized.  From
(\ref{ComplexAmplitude}) we can see that, when $\Phi = \Psi$,
\begin{eqnarray}\label{Varrho}
\varrho = \rho v^{0},
\end{eqnarray}
where $\rho$ is also from the wavefunction $\psi$.  We define the
angular velocities as
\begin{eqnarray}
\Omega_{\mu}  & :=  & 2(\partial_{\mu}R)\widetilde{R} \label{G5a1} \\
\omega_{\mu} & := & 2(\partial_{\mu}R_{0})\widetilde{R}_{0}.
                \label{G5a2}
\end{eqnarray}
Because $R\widetilde{R} = 1$, we have that
$(\partial_{\mu}R)\widetilde{R} = - R(\partial_{\mu}
\widetilde{R})$.  Notice that this implies that $\Omega_{\mu} = -
\widetilde{\Omega}_{\mu}$.  Similarly, we can show that
$\omega_{\mu} = - \widetilde{\omega}_{\mu}$. Now, the spin
polarization vector $s = s^{\mu}\gam{\mu}$ is given by
\begin{eqnarray}\label{G7}
s & := & \frac{1}{2}R\gamma_{3}\widetilde{R} \\
    & = & \frac{1}{2}R_{0}\gamma_{3}\widetilde{R}_{0}. \label{sEq2}
\end{eqnarray}
Finally, the spin angular momentum bivector $S = \I s v$ is
defined by
\begin{eqnarray}
S & := & \frac{1}{2} R \I \sigma_{3} \widetilde{R} \label{G7a}\\
    & = & \frac{1}{2} R_{0} \I \sigma_{3} \widetilde{R}_{0}.
    \label{SEq2}
\end{eqnarray}
Equation (\ref{G7a}) gives us $\partial_{\mu} S = 1/2
(\Omega_{\mu} S - S \Omega_{\mu})$.  Similarly, from (\ref{SEq2}),
we have $\partial_{\mu} S = 1/2 (\omega_{\mu} S - S
\omega_{\mu})$. From (\ref{vEq2}), (\ref{sEq2}) and (\ref{SEq2})
we see that $R_{0}$ determines the orientations of $v$, $s$ and
$S$. By orientation we mean the directions of the vectors $v$ and
$s$, and the ``tilt'' and direction of the directed area $S$, in
the four dimensional spacetime.\newline\newline

%----------------BEGIN EQUATION DERIVATIONS---------------------------------------
%\section{PHASE FORMULAS}
%\label{Formulas}
\noindent\textbf{3. PHASE FORMULAS}\newline

Here we derive the spacetime algebra formulas for the dynamic and
geometric phases.  Initially we  simply translate the standard
dynamic phase formula into Hestenes' formalism.  But, much of
Hestenes' theory deals with local observables [\ref{H73}], so we
will define a local dynamic phase.  This local definition is
useful for working in the spacetime algebra.  Next we derive the
geometric phase formula.  We then briefly examine the
relationships between the spacetime algebra and standard formulas.
Finally, we will find that we are able to redefine the phases,
allowing us to use much simpler formulas in Hestenes' theory.

First, notice that (\ref{DiracEquation}) is invariant under the
transformation
\begin{eqnarray}
\psi~\rightarrow~\psi~\exp{[\I\sigma_{3}\alpha]},
\end{eqnarray}
where $\alpha$ is a real constant.  In the Dirac theory, this
corresponds to adding a constant phase to $\Psi$.  In Hestenes'
theory, we are adding a constant angle to the Euler angle $\chi$
in (\ref{DiracSpinor2}).  Let us now define the projection
operator $\Pi$ by
\begin{eqnarray}\label{Projection}
\Pi(\psi) & := & \left\{\psi^{'} : \psi^{'} = \psi
            e^{\I\sigma_{3}\alpha}\mbox{,}\ \mbox{for all real
            constants}\ \alpha \right\}
\end{eqnarray}
and let $\mathcal{P}$ be the space of all such projections.  This
amounts to projecting the multivectors $\psi$ in the Hilbert space
$\mathcal{H}$ onto the representative ray $\Pi(\psi)$ in
$\mathcal{P}$, where $\Pi(\psi)$ differs from $\psi$ only by a
constant spatial rotation $\exp[\I\sig{3}\alpha]$.  [Equation
(\ref{Projection}) corresponds to the projection of $\Psi$ onto
the ray $\widehat{\Pi}(\Psi)$, where $\widehat{\Pi}(\Psi)$ differs
from $\Psi$ only by a phase factor $\exp[i\alpha]$.]  If
$\psi(\xi)$ evolves along the curve $\mathcal{C}$ in
$\mathcal{H}$, then $\Pi(\psi(\xi))$ will evolve along the curve
$\widehat{\mathcal{C}}$ in $\mathcal{P}$.  Notice that we have
parameterized $\psi$'s evolution in $\mathcal{H}$ by $\xi$.  The
geometric phase should only depend on the curve
$\widehat{\mathcal{C}}$ in $\mathcal{P}$, the path of $\Pi(\psi)$
in $\mathcal{P}$.  In particular, the geometric phase needs to be
independent of the rate at which $\Pi(\psi)$ traverses
$\widehat{\mathcal{C}}$.  Therefore, a reparameterization of $\xi$
can not affect the geometric phase.

The [complex] global dynamic phase, $\Delta_{G}$, is given by
[\ref{MuS1}]
\begin{eqnarray}\label{GlobalDP}
\Delta_{G} & = & \int d\xi d^{3}\! x\ \Im(\Psi^{\dag}\dot{\Psi)},
\end{eqnarray}
where the overdot represents differentiation with respect to
$\xi$.  As we have seen earlier, $\Im(\Psi^{\dag}\dot{\Psi)}
\leftrightarrow -\scalar{\dot{\psi}\I\sig{3} \gam{0}
\widetilde{\psi}\gam{0}}$.  So the spacetime algebra formula for
the global dynamic phase, $\delta_{G}$, is
\begin{eqnarray}\label{STAGlobalDP}
\delta_{G} & = & - \int d\xi d^{3}\! x\
            \scalar{\dot{\psi}\I\sig{3}
            \gam{0} \widetilde{\psi}\gam{0}}.
\end{eqnarray}
From (\ref{STAGlobalDP}) let us define the local dynamic phase by
\begin{eqnarray}\label{LocalDP}
\varrho\ \dot{\delta}_{L} & := & -\scalar{\dot{\psi}\I\sig{3}
            \gam{0} \widetilde{\psi}\gam{0}}.
\end{eqnarray}
We use $\varrho$, rather than $\rho$, in (\ref{LocalDP}) because
the Hermitian adjoint used in (\ref{STAGlobalDP}) is frame
dependent, so we expect the same for our probability distribution.
Also, since the Hermitian adjoint singles out a preferred time
direction, we will let $\xi$ be the \textit{observer's time},
given by $t$. Using (\ref{DiracSpinor}) in (\ref{LocalDP}) results
in, after some tedious algebra,
\begin{eqnarray}\label{LocalDP2}
\varrho\ \dot{\delta}_{L} & = & \rho \dot{\beta}
                    \scalar{\frac{1}{2}\ R \gam{3} \widetilde{R}
                    \gam{0}} - \rho \scalar{\dot{R} \I\gam{3}
                    \widetilde{R} \gam{0}}  \nonumber       \\
                    & = & \rho \dot{\beta} \scalar{s\gam{0}} -
                    \rho \scalar{\Omega_{0} \I s \gam{0}}
                    \nonumber               \\
                    & = & \rho s^{0} \dot{\beta} - \rho \scalar{
                    \Omega_{0} S v \gam{0}} \nonumber       \\
                    & = & \varrho\ \frac{s^{0}}{v^{0}}\ \dot{\beta}
                    - \rho \scalar{\Omega_{0} S (v^{0} +
                    \mathbf{v})},
\end{eqnarray}
where $v\gam{0} := v^{0} + \mathbf{v}$, $\mathbf{v} :=
v^{n}\sig{n}$, and $S v = \I s$ because $vv =1$.  The $\Omega_{0}
S$ term in (\ref{LocalDP2}) can have scalar, bivector and
pseudoscalar [i.e., \I] parts, denoted by $\Omega_{0} \cdot S$,
$\multi{\Omega_{0}S}{2}$ and $\multi{\Omega_{0}S}{4}$,
respectively. Under the reversion operation, the scalar and
pseudoscalar parts are even, while the bivector part is odd.
Hence,
\begin{eqnarray}\label{star}\
\multi{\Omega_{0}S}{2} & = & \frac{1}{2}\left(\Omega_{0} S -
                       ( \widetilde{\Omega_{0} S})\right)
                        \nonumber           \\
                    & = & \frac{1}{2}(\Omega_{0} S - S
                            \Omega_{0}).
\end{eqnarray}
From (\ref{G7a}), we see that
\begin{eqnarray}\label{Sdot}
\dot{S} & = & \frac{1}{2}(\dot{R} \I \sig{3} \widetilde{R} + R \I
                \sig{3} \dot{\widetilde{R}})    \nonumber   \\
        & = & \multi{\Omega_{0}S}{2}.
\end{eqnarray}
Using this in (\ref{LocalDP2}) gives us
\begin{eqnarray}\label{LDP3}
\varrho\ \dot{\delta}_{L} & = & \varrho\ \frac{s^{0}}{v^{0}}
        \dot{\beta} - \rho \scalar{(\Omega_{0} \cdot S + \dot{S}
         + \multi{\Omega_{0}S}{4})(v^{0} + \mathbf{v})}
        \nonumber                                       \\
        & = & \varrho\ \frac{s^{0}}{v^{0}} \dot{\beta} - \varrho
        \Omega_{0} \cdot S - \rho \scalar{\dot{S} \mathbf{v}}.
\end{eqnarray}

Now, $s \cdot v = 0$, so
\begin{eqnarray} \label{s0}
s^{0} & = & \frac{\mathbf{s} \cdot \mathbf{v}}{v^{0}},
\end{eqnarray}
where $s\gam{0} = s^{0} + \mathbf{s}$.  Thus, (\ref{LDP3}) becomes
\begin{eqnarray}
\dot{\delta}_{L} & = & -\Omega_{0}\cdot S -
            \frac{\mathbf{v}}{v^{0}} \cdot \left[
            \dot{S} - \frac{\mathbf{s}}{v^{0}} \dot{\beta}
                \right].  \label{Delta}
\end{eqnarray}
From (\ref{STAGlobalDP}) to (\ref{Delta}), it follows that, after
letting $\xi = t$,
\begin{eqnarray}\label{STAGlobalDP2}
\delta_{G} & = & \int dt d^{3}\!x\ \varrho \dot{\delta}_{L}
                \nonumber           \\
            & = & \int dt d^{3}\!x\ \varrho \left\{ -\Omega_{0}\cdot S
             - \frac{\mathbf{v}}{v^{0}} \cdot \left[
            \dot{S} - \frac{\mathbf{s}}{v^{0}} \dot{\beta} \right]
            \right\}.
\end{eqnarray}

We now find a spacetime algebra geometric phase formula. First,
let us define a new wavefunction $\psi^{'} :=
\psi\exp[-\I\sig{3}\delta_{L}]$ that differs from $\psi$ only in
having the local dynamic phase removed.  Now,
\begin{eqnarray}\label{PsiPrime}
\dot{\psi}^{'} & = & \dot{\psi}e^{-\I\sig{3}\delta_{L}} + \psi
        e^{-\I\sig{3}\delta_{L}}\left(-\I\sig{3}\dot{\delta}_{L}
        \right).
\end{eqnarray}
Using (\ref{ComplexAmplitude}) for the inner product of
$\dot{\psi}^{'}$ with $\psi^{'}$, and noting that $\scalar{\psi
\I\sig{3} \gam{0} \widetilde{\psi} \gam{0}} = 0$, we have from
(\ref{PsiPrime})
\begin{eqnarray}\label{NewEqn}
\scalar{\dot{\psi}^{'} \gam{0} \widetilde{\psi}^{'} \gam{0}} -
\I\sig{3} \scalar{\dot{\psi}^{'} \I\sig{3} \gam{0}
\widetilde{\psi}^{'} \gam{0}} & = & \scalar{\dot{\psi} \gam{0}
    \widetilde{\psi} \gam{0}} - \I\sig{3} \scalar{\dot{\psi}
    \I\sig{3} \gam{0} \widetilde{\psi} \gam{0}}    \nonumber    \\
    & & - \I\sig{3}
    \dot{\delta}_{L} \scalar{ \psi \gam{0} \widetilde{\psi}
    \gam{0}}        \nonumber       \\
    & = & \scalar{\dot{\psi} \gam{0}
    \widetilde{\psi} \gam{0}} + \I\sig{3} \varrho \dot{\delta}_{L}
    - \I\sig{3} \varrho \dot{\delta}_{L}    \nonumber       \\
    & = & \scalar{\dot{\psi} \gam{0} \widetilde{\psi} \gam{0}}.
\end{eqnarray}
Since $\scalar{\dot{\psi}^{'} \gam{0} \widetilde{\psi}^{'}
\gam{0}} = \scalar{\dot{\psi} \gam{0} \widetilde{\psi} \gam{0}}$,
it follows from (\ref{NewEqn}) that
\begin{eqnarray}\label{ZeroEqn}
\scalar{\dot{\psi}^{'} \I\sig{3} \gam{0} \widetilde{\psi}^{'}
            \gam{0}} & = & 0.
\end{eqnarray}

With $\psi^{'}$ differing from $\psi$ only by the $\chi$ factor of
(\ref{DiracSpinor2}), let us write
\begin{eqnarray}\label{PsiPrime2}
\psi^{'} & = & \left( \rho e^{\I\beta}\right)^{1/2} L U_{0}
        e^{-\I\sig{3} \chi^{'}/2}.
\end{eqnarray}
Using (\ref{PsiPrime2}) in (\ref{ZeroEqn}), and doing some lengthy
algebra, gives us, similar to our derivation of (\ref{LDP3}),
\begin{eqnarray}\label{ChiEqn1}
\scalar{\dot{\psi}^{'} \I\sig{3} \gam{0} \widetilde{\psi}^{'}
        \gam{0}} & = & \frac{\rho}{2} \scalar{(- R_{0} \I\sig{3}
        \dot{\chi}^{'} + \dot{\beta} \I
        R_{0} + 2 \dot{R}_{0} )\I\sig{3} \gam{0} \widetilde{R}_{0}
        \gam{0}}    \nonumber       \\
        & = & \frac{\varrho}{2} \dot{\chi}^{'} - \varrho
        \frac{s^{0}}{v^{0}} \dot{\beta} + \rho \scalar{ \omega_{0}
        S (v^{0} + \mathbf{v})} \nonumber       \\
        & = & \frac{\varrho}{2} \dot{\chi}^{'} - \varrho
        \frac{s^{0}}{v^{0}} \dot{\beta} + \varrho \omega_{0} \cdot S +
        \rho \scalar{\dot{S} \mathbf{v}}    \nonumber       \\
        & = & \frac{\varrho}{2} \dot{\chi}^{'} - \varrho
        \frac{s^{0}}{v^{0}} \dot{\beta} + \varrho \omega_{0} \cdot
        S + \varrho \frac{\mathbf{v}}{v^{0}} \cdot \dot{S},
\end{eqnarray}
where we used the fact that $\dot{S} = \multi{\omega_{0} S}{2}$
[see (\ref{SEq2})].  Then, from (\ref{ZeroEqn}) and
(\ref{ChiEqn1}), we have
\begin{eqnarray}
\frac{1}{2}\ \dot{\chi}^{'} & = & - \omega_{0} \cdot S -
        \frac{\mathbf{v}}{v^{0}} \cdot \left[ \dot{S} -
        \frac{\mathbf{s}}{v^{0}}  \dot{\beta} \right].   \label{ChiEqn3}
\end{eqnarray}

Notice that all of the terms on the righthand-side of
(\ref{ChiEqn3}) depend only on the path $\widehat{\mathcal{C}}$ in
$\mathcal{P}$. That is, they are determined by $R_{0}$, not by
$R$.  [This does not hold for $\delta_{L}$ because of the presence
of $\Omega_{0}$ in (\ref{Delta}).]  Also, both sides of
(\ref{ChiEqn3}) are linear in the time derivative.  Hence,
$\chi^{'}$ is independent of a reparameterization of $t$.  These
two observations allow us to define the local spacetime algebra
geometric phase by $\gam{L} := - \chi^{'}/2 + \chi(0)/2$.  [We use
a minus sign here because $-\chi^{'}/2 = -\chi/2 - \delta_{L}$.
The $\chi(0)/2$ factor is because $\gam{L}(0)$ must vanish. Then
$\gamma_{L} + \delta_{L} = -\chi/2 + \chi(0)/2$, where we have
previously identified $-\chi/2$ as the total phase.] Thus,
\begin{eqnarray}
\dot{\delta}_{L} & =  & -\Omega_{0}\cdot S -
            \frac{\mathbf{v}}{v^{0}} \cdot \left[
            \dot{S} - \frac{\mathbf{s}}{v^{0}} \dot{\beta} \right] \label{LDP} \\
\dot{\gamma}_{L} & = &  \omega_{0} \cdot S +
            \frac{\mathbf{v}}{v^{0}} \cdot \left[ \dot{S} -
           \frac{\mathbf{s}}{v^{0}}  \dot{\beta} \right],
             \label{LGP}
\end{eqnarray}
where $\delta_{L}(0) = 0$ and $\gamma_{L}(0) = 0$.  To find the
global spacetime algebra geometric phase, $\gam{G}$, we need to
perform the integration
\begin{eqnarray}\label{GlobalGP}
\gam{G} & := & \int dt d^{3}\! x\ \varrho \dot{\gamma}_{L}
                \nonumber           \\
        & = & \int dt d^{3}\!x\ \varrho \left\{  \omega_{0}
        \cdot S +  \frac{\mathbf{v}}{v^{0}} \cdot \left[ \dot{S} -
        \frac{\mathbf{s}}{v^{0}}  \dot{\beta} \right]  \right\},
\end{eqnarray}
similar to (\ref{STAGlobalDP2}) above.

Equation (\ref{LDP}) is a straightforward translation of the
standard dynamic phase formula into the spacetime algebra.  Thus,
the value of $\delta_{G}$ will equal that of $\Delta_{G}$.  We may
ask if the same holds true for $\gam{G}$ and the standard
geometric phase $\Gamma_{G}$?  This will not generally be the
case.  The $-\chi/2$ factor in (\ref{DiracSpinor2}), which we
identified as the total phase, may contain dynamics beyond the
local representations of $\Delta_{G}$ and $\Gamma_{G}$.  As an
example, let us consider an adiabatic evolution of a wavefunction
that starts in an energy eigenstate.  Let the $m$-th energy
eigenfunction be given by
\begin{eqnarray}\label{energyeigen}
\phi_{m} & = & \left(\rho e^{\I\beta}\right)^{1/2} R_{m0}
            e^{-\I\sig{3} \chi_{m}/2}.
\end{eqnarray}
The adiabatic theorem states that our wavefunction $\psi_{m}$ can
differ from $\phi_{m}$ only by a phase factor.  Thus, using the
complex global phases $\Delta_{mG}$ and $\Gamma_{mG}$
\begin{eqnarray}\label{psim}
\psi_{m} & = & \phi_{m} e^{\I\sig{3}(\Delta_{mG} + \Gamma_{mG})}.
\end{eqnarray}
[Strictly speaking, (\ref{psim}) will contain some
$\mathcal{O}(\varepsilon)$ terms where $\varepsilon$ is an
infinitesimal number that reflects the degree of adiabaticity, and
the $\mathcal{O}(\varepsilon)$ terms are due to departures from
strict adiabaticity.  We will ignore these terms in (\ref{psim}).]
In this situation, $\Gamma_{mG}$ is given by [\ref{WaLi99}]
\begin{eqnarray}\label{GPadaibatic}
\dot{\Gamma}_{mG} & = & -\int d^{3}\!x \Im \left(\Phi_{m}^{\dag}
            \dot{\Phi}_{m} \right).
\end{eqnarray}
A derivation similar to that in Section 3 for the dynamic phase
shows that
\begin{eqnarray}\label{GPadiabatic2}
\dot{\Gamma}_{mG} & = & \int d^{3}\!x \left(\dot{\gamma}_{mL} +
            \frac{\dot{\chi}}{2}\right)     \nonumber       \\
            & = & \dot{\gamma}_{mG} + \int d^{3}\!x
            \frac{\dot{\chi}}{2}.
\end{eqnarray}
This result does not invalidate calling $\gamma_{L}$ the spacetime
algebra geometric phase.  It still represents that part of the
total phase $-\chi/2$ due only to the geometry of the
wavefunction's evolution.

Now, (\ref{DiracSpinor}) can be rewritten as
\begin{eqnarray}\label{minus}
\psi & = & \left(\rho e^{\I\beta}\right)^{1/2} R_{0} e^{\I\sig{3}
        (\gam{L} + \delta_{L} - \chi(0)/2)} \nonumber       \\
        & = & \left(\rho e^{\I\beta}\right)^{1/2} R_{0} e^{\I\sig{3}
        (\int^{t}d\tau(\omega_{0}\cdot S - \Omega_{0}\cdot S) -
        \chi(0)/2)}.
\end{eqnarray}
Thus, we can redefine the phases as
\begin{eqnarray}
\dot{\widehat{\delta}}_{L} & := & -\Omega_{0}\cdot S  \label{twominus}        \\
\dot{\widehat{\gamma}}_{L} & := & \omega_{0} \cdot S,
                \label{threeminus}
\end{eqnarray}
where $\widehat{\delta}_{L}(0) = 0$ and $\widehat{\gamma}_{L}(0) =
0$.  Equations (\ref{twominus}) and (\ref{threeminus}) are still
physically meaningful definitions. They capture all the essential
properties of the dynamic and geometric phases.  The simplicity of
(\ref{twominus}) and (\ref{threeminus}), versus (\ref{LDP}) and
(\ref{LGP}), respectively, suggests that they are more useful
definitions for Hestenes' formalism.  It is easy to show, for
example, that (\ref{twominus}) and (\ref{threeminus}) hold exactly
in the non-relativistic limit.  That is, the second terms on the
right-hand sides of (\ref{LDP}) and (\ref{LGP}) vanish in the
non-relativistic limit because $\mathbf{v} \rightarrow
\mathbf{0}$. Thus, we can use exactly the same phase formulas in
the relativistic and non-relativistic cases.  Also, it is easy to
show the (\ref{LDP}) and (\ref{LGP}) reduce to (\ref{twominus})
and (\ref{threeminus}), respectively, when we consider the
adiabatic evolution of an energy eigenstate.  Overall,
(\ref{twominus}) and (\ref{threeminus}) seem to be more reasonable
definitions of the phases, in the spacetime algebra, than
(\ref{LDP}) and (\ref{LGP}).  Finally, we note that Hestenes' had
previously proposed using $\Omega_{0} \cdot S$ as the total local
phase formula [\ref{H93}]. As we see, except for the sign, this is
only the dynamic part of the phase. We also need to take into
account the geometric part of the total local phase.

The formulas (\ref{twominus}) and (\ref{threeminus}) can be
written as
\begin{eqnarray}
\dot{\widehat{\delta}}_{L} & = & -\scalar{\dot{R} \I\sig{3}
            \widetilde{R}}      \label{RLDP}        \\
\dot{\widehat{\gamma}}_{L} & = & \scalar{\dot{R}_{0} \I\sig{3}
            \widetilde{R}_{0}} .     \label{RLGP}
\end{eqnarray}
Unlike (\ref{HermitianAdjoint}) and (\ref{ComplexAmplitude}),
these no longer have $\gam{0}$ present.  Thus, they are frame
independent quantities.  Now, the wavefunction $\psi$ determines a
set of \textit{streamlines} via the proper velocity vector $v$.
That is, if we imagine a particle as actually starting at some
initial spacetime point $x_{0}$, the spacetime vector $v$ will
determine its future positions, given by the streamlines.  Along a
given streamline there is a proper time $\tau$.  Because
(\ref{RLDP}) and (\ref{RLGP}) are independent of a particular
reference frame, we can allow the derivative to be with respect to
$\tau$.  Hence, (\ref{RLDP}) and (\ref{RLGP}) can also be used for
the proper phase formulas [remembering that, in this case, we are
using a proper time derivative].\newline\newline

%-------------BEGIN DISCUSSION-----------------------------
%\section{DISCUSSION}
%\label{Discussion}
\noindent\textbf{4. DISCUSSION}\newline

Let us first review a few facts about $\beta$ that will be useful
in the following discussion [see the article by Gull, Lasenby and
Doran [\ref{GLD2}] for more background].  When there is no
electromagnetic field, i.e., $A= 0$ in (\ref{DiracEquation}), the
Dirac equation admits plane wave solutions.  For the electron
solutions $\beta = 0$, while for the positron solutions $\beta =
\pi$.  However, when $A \neq 0$, a general wavefunction can have
other values of $\beta$, as demonstrated by the [non-relativistic]
solutions for the hydrogen atom.  Additionally, in their numerical
simulations of tunnelling times, Gull et al.\ show that $\rho$ and
$\beta$ are not necessarily constant at a given position for all
time.  So the $\dot{\beta}$ term in (\ref{LDP}) does not
necessarily vanish.

Now, it can be shown that [\ref{DLG}]
\begin{eqnarray}
\widehat{\gamma}_{5} \Psi & \leftrightarrow & \psi \sig{3}
                \label{1}                               \\
\overline{\Psi}\Psi & \leftrightarrow & \rho \cos(\beta)
                \label{2}                               \\
\overline{\Psi}i\widehat{\gamma}_{5} \Psi & \leftrightarrow &
                -\rho \sin(\beta),     \label{3}
\end{eqnarray}
where $\widehat{\gamma}_{5}$ is the standard Dirac matrix.  It
follows that
\begin{eqnarray}\label{chiralbeta}
\overline{\Psi}\Psi - i\widehat{\gamma}_{5}\overline{\Psi}
                i\widehat{\gamma}_{5}
                \Psi & = & \rho e^{i\hat{\gamma}_{5}\beta}
                \nonumber                       \\
                & \leftrightarrow & \rho e^{\I\beta}.
\end{eqnarray}
So the $e^{\I\beta}$ term can be thought of as a local
\textit{chiral transformation} [\ref{Gros93}], an observation made
previously [\ref{H82}].  Let us take this literally and think of
$\beta$ as the \textit{chiral angle} in the Dirac wavefunction.

Equation (\ref{LDP}) for the dynamic phase is a straightforward
translation of the standard phase formula (\ref{GlobalDP}) into
the geometric algebra.  We can rewrite (\ref{GlobalDP}) as
\begin{eqnarray}\label{GPdot}
\dot{\Delta}_{G} & = & \int d^{3}\! x\ \Im(\Psi^{\dag} \dot{\Psi})
              \nonumber  \\
        & = & - \int d^{3}\! x\ \Psi^{\dag}i\frac{d}{dt} \Psi.
\end{eqnarray}
In the standard Dirac theory, (\ref{GPdot}) is interpreted as the
negative of the expected value of the energy operator.  The same
physical interpretation is then given to (\ref{LDP}), locally.
However, interpreting (\ref{LDP}) using only the spacetime algebra
formalism is difficult because of the presence of $\beta$.

In contrast, the dynamic phase formula in (\ref{twominus}) is
easily interpreted in the spacetime algebra.  It is the negative
of the component of $\Omega_{0}$ in the spacetime plane $S$.  To
translate (\ref{twominus}) back into the standard Dirac theory, we
use (\ref{RLDP}) and (\ref{ComplexAmplitude2}).  If $R$ is the
spacetime representation of $\mathbf{R}$, in the standard theory
(\ref{RLDP}) is given by
\begin{eqnarray}\label{DeltaL1}
\dot{\widehat{\Delta}}_{L} & = & \Im(\overline{\mathbf{R}}
        \dot{\mathbf{R}}).
\end{eqnarray}
Because $\widetilde{R}R = 1$, we have from (\ref{chiralbeta}) that
$\overline{\mathbf{R}}\mathbf{R} = 1$.  It follows that
$\overline{\mathbf{R}}\dot{\mathbf{R}} = - (\overline{\mathbf{R}}
\dot{\mathbf{R}})^{\dag}$.  Hence, $\overline{\mathbf{R}}
\dot{\mathbf{R}}$ is purely imaginary.  So (\ref{DeltaL1}) becomes
\begin{eqnarray}\label{DeltaL2}
\dot{\widehat{\Delta}}_{L} = -i \overline{\mathbf{R}}
        \dot{\mathbf{R}}.
\end{eqnarray}
The question now is, how to find $\mathbf{R}$?  Since
\begin{eqnarray}\label{igamma5}
i\widehat{\gamma}_{5} \Psi & \leftrightarrow & \I \psi
\end{eqnarray}
we can write (\ref{DiracSpinor}) in the standard theory as
\begin{eqnarray}\label{PsiStandard}
\Psi & = & \sqrt{\rho} e^{i\hat{\gamma}_{5} \beta /2}
            \mathbf{R}.
\end{eqnarray}
Provided $\rho \neq 0$, we have that
\begin{eqnarray}\label{mathbfR}
\mathbf{R} & = & \frac{e^{-i\hat{\gamma}_{5} \beta
        /2}}{\sqrt{\rho}} \Psi.
\end{eqnarray}
So $\mathbf{R}$ is given by a local chiral transformation of
$\Psi$.  Notice that (\ref{PsiStandard}) implies that $\mathbf{R}$
is invariant under a local chiral transformation of $\Psi$.

For both sets of phase formulas, we see that they are easily
interpreted in one formalism but not the other.  Both geometric
phase definitions are invariant under a local gauge
transformation.  [This likely accounts for the difference in
$\gamma_{G}$ and $\Gamma_{G}$, see (\ref{GPadiabatic2}).]  The
phases in (\ref{twominus}) and (\ref{threeminus}) are also
invariant under a local chiral transformation.  These may prove to
be more useful in the electroweak theory.

It is difficult to decide theoretically which set of formulas is
the correct one.  It may also be difficult to experimentally
verify which set is correct.  This is because (\ref{LDP}) and
(\ref{LGP}) reduce to (\ref{twominus}) and (\ref{threeminus}) for
adiabatic evolutions of energy eigenstates and, in the
non-relativistic limit.\newline\newline

%--------------BEGIN ACKNOWLEDGEMENTS-----------------------
%\section{ACKNOWLEDGEMENTS}
%\label{Acknowledgements}
\noindent\textbf{ACKNOWLEDGEMENTS}\newline

DWD and PMY would like to thank the NSF for grant \#9732986.  DWD
would like to thank Dr.\ Andr\'{e} Gsponer for his invaluable
correspondence regarding this paper. The authors also thank the
reviewers for their very helpful comments.\newline\newline

\noindent\textbf{REFERENCES} \newline
\begin{enumerate}
\item\label{DLG} C. Doran, A. Lasenby and S. Gull, ``States and
        operators in the spacetime algebra,'' \textit{Found. Phys.}
        \textbf{23}, 1239-1264 (1993).\vspace*{-.1in}
\item\label{Gros93} F. Gross, \textit{Relativistic Quantum Mechanics and Field
        Theory} (Wiley Interscience, New York, 1993).\vspace*{-.1in}
\item\label{Gspon2002} A. Gsponer, ``On the ``equivalence'' of the {M}axwell
        and {D}irac equations,'' \textit{Int. J. Theor. Phys.}
        \textbf{41}, 689-694 (2002).\vspace*{-.1in}
\item\label{GLD1} S. Gull, A. Lasenby and C. Doran, ``Imaginary numbers are not
        real: The geometric algebra of spacetime,''
        \textit{Found. Phys.} \textbf{23}, 1175-1201 (1993).\vspace*{-.1in}
\item\label{GLD2} S. Gull, A. Lasenby and C. Doran, ``Electron paths, tunnelling and diffraction
        in the spacetime algebra,'' \textit{Found. Phys.}
        \textbf{23}, 1329-1356 (1993).\vspace*{-.1in}
\item\label{HSTA} D. Hestenes, \textit{Space-Time Algebra} (Gordon
        \& Breach, New York, 1966).\vspace*{-.1in}
\item\label{H73} D. Hestenes, ``Local observables in quantum
        theory,'' \textit{J. Math. Phys. (N.Y.)} \textbf{14}, 893-905
        (1973).\vspace*{-.1in}
\item\label{H75-1} D. Hestenes, ``Observables, operators and complex numbers
        in the {D}irac theory,'' \textit{J. Math. Phys. (N.Y.)}
        \textbf{16}, 556-572 (1975).\vspace*{-.1in}
\item\label{H82} D. Hestenes, ``Spacetime structure of weak and electromagnetic
        interactions,'' \textit{Found. Phys.} \textbf{12}, 153-168
        (1982).\vspace*{-.1in}
\item\label{H93} D. Hestenes, ``Zitterbewegung modeling,''
        \textit{Found. Phys.} \textbf{23}, 365-387 (1993).\vspace*{-.1in}
\item\label{MuS1} N. Mukunda and R. Simon, ``Quantum kinematic approach to
        the geometric phase {I}: General formalism,'' \textit{Ann.
        Phys. (N.Y.)} \textbf{228}, 205-268 (1993).\vspace*{-.1in}
\item\label{Sak} J.J. Sakurai, \textit{Modern Quantum Mechanics}
        (Addison-Wesley, Reading, MA, 1994).\vspace*{-.1in}
\item\label{WaLi99} Z. Wang and B. Li, ``Geometric phase in relativistic quantum
        mechanics,'' \textit{Phys. Rev. A} \textbf{60}, 4313-4317
        (1999).
\end{enumerate}

\end{document}